\documentclass[prl,twocolumn,showpacs,superscriptaddress,floatfix]{revtex4}
\usepackage{graphicx,amsfonts,amssymb,amsmath, hyperref}

\newif\ifhyper
\hypertrue
\ifhyper
\hypersetup{
   citecolor = {green},
   colorlinks = {true}, 
   urlcolor = {blue} 
}
\fi

\newcommand{\rmi}{\mathrm{i}}

\newcommand{\anb}{b^{\phantom\dagger}}
\newcommand{\crb}{b^\dagger}
\newcommand{\bn}{{\boldsymbol{n}}}
\newcommand{\bi}{{\boldsymbol{i}}}
\newcommand{\bj}{{\boldsymbol{j}}}
\newcommand{\bk}{{\boldsymbol{k}}}
\newcommand{\bl}{{\boldsymbol{l}}}

\begin{document}

\title{Emergent fermions and anyons in the Kitaev model}

\author{Kai Phillip Schmidt}
\email{kaiphillip.schmidt@epfl.ch}
\affiliation{Institute of Theoretical Physics, \'{E}cole Polytechnique
 F\'{e}d\'{e}rale de Lausanne, CH-1015 Lausanne, Switzerland}
 
\author{S\'ebastien Dusuel}
\email{sdusuel@gmail.com}
\affiliation{Lyc\'ee Louis Thuillier, 70 Boulevard de Saint Quentin,
  80098 Amiens Cedex 3, France}

\author{Julien Vidal}
\email{vidal@lptmc.jussieu.fr}
\affiliation{Laboratoire de Physique Th\'eorique de la Mati\`ere Condens\'ee,
  CNRS UMR 7600, Universit\'e Pierre et Marie Curie, 4 Place Jussieu, 75252
  Paris Cedex 05, France}


\begin{abstract}

We study the gapped phase of the Kitaev model on the honeycomb lattice using
perturbative continuous unitary transformations. The effective low-energy
Hamiltonian is found to be an extended toric code with interacting
anyons. High-energy excitations are emerging free fermions which are composed
of hardcore bosons with an attached string of spin operators. The excitation
spectrum is mapped onto that of a single particle hopping on a square lattice
in a magnetic field. We also illustrate how to compute correlation functions in this framework. 
The present approach yields analytical perturbative
results in the thermodynamical limit without using the Majorana or the Jordan-Wigner fermionization
initially proposed to solve this problem.
\end{abstract}


\maketitle


The study of elementary excitations in strongly correlated systems is a
fascinating field of current research. As exemplified in the fractional quantum
Hall effect, such excitations can be very different from the elementary
constituents present in the system.
In the same spirit, emergent fermions and gauge fields in purely boson/spin
systems have attracted much attention recently \cite{Wen04}. The emergence of
anyonic excitations in two-dimensional systems has also triggered a tremendous
amount of interest, especially since its relevance for topological quantum
computation (see Ref.~\onlinecite{DasSarma07} for a recent review) has been
pointed out by Kitaev in a seminal paper \cite{Kitaev03} introducing the
celebrated toric code.

\begin{figure}[ht]
  \includegraphics[width=\columnwidth]{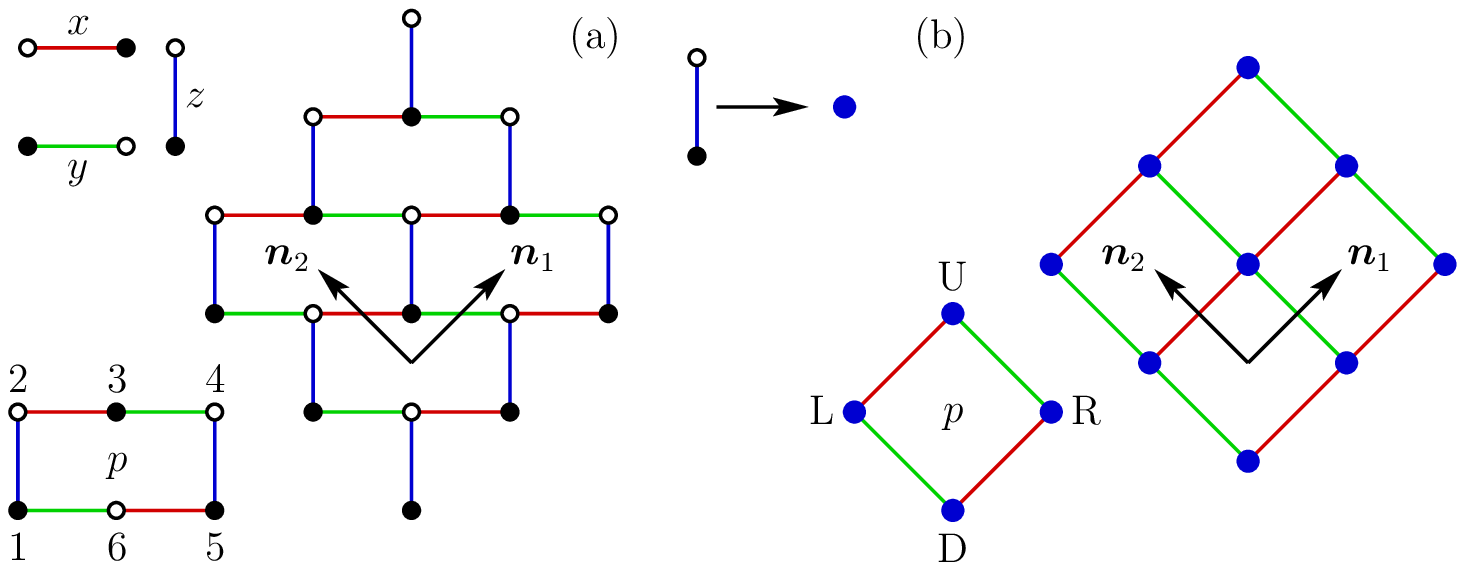}
  \caption{(color online).
    Mapping of the brick-wall lattice (a) to an effective square
    lattice (b) with unit basis vectors $\bn_1$ and $\bn_2$. The numbering of
    the sites of a plaquette $p$ is shown in both cases.}
  \label{fig:mapping_brickwall_square}
\end{figure}

More recently, Kitaev introduced a more realistic model \cite{Kitaev06} for which experiments using ultracold atoms or polar molecules have been proposed \cite{Duan03,Micheli06}.
This model is a two-dimensional spin-$1/2$ system on the honeycomb or brick-wall lattice, as illustrated in Fig.~\ref{fig:mapping_brickwall_square}. 
It consists solely of Ising-like interactions but in different quantization axis. More precisely, the Hamiltonian reads
%
%
\begin{equation}
  \label{eq:ham}
  H=-\sum_{\alpha=x,y,z}\sum_{\alpha-\mathrm{links}}
  J_\alpha\,\sigma_i^\alpha\sigma_j^\alpha,
\end{equation} 
%
%
where $\sigma_i^\alpha$ are the usual Pauli matrices at site $i$. Without loss of generality \cite{Kitaev06}, in the following, we assume  $J_\alpha \geq 0$ for all $\alpha$ and $J_z\geq J_x,J_y$. 

Kitaev solved the model exactly by introducing Majorana fermions to represent the spin
operators in an extended Hilbert space. In this way, the Hamiltonian is reduced
to free fermions in a static $\mathbb{Z}_2$ gauge field on the honeycomb lattice. The physical states are selected by a projection step which simply amounts to a selection rule on the parity of the number of fermions \cite{Vidal08_2}, in agreement with the general conclusions of Ref.~\onlinecite{Levin03}.
The system exhibits a gapless phase for $J_x+J_y>J_z$, with non-Abelian anyonic excitations arising when a magnetic field is switched on \cite{Kitaev06}. {\it A contrario},  for $J_x+J_y<J_z$, the system is gapped and the low-energy effective Hamiltonian,  at lowest nontrivial order in perturbation ($J_x,J_y \ll J_z$) and periodic boundary conditions, turns out to be exactly the toric code. Consequently, at this order, free Abelian anyons are present in the gapped phase.

%
%
%
%
Note also that an alternative treatment has been proposed \cite{Chen07_1,Chen07_2,Feng07} based on the Jordan-Wigner transformation which transforms the spin system into a system of fermions with $p$-wave BCS pairing and a site-dependent chemical potential. 
%
%

In this Letter, we focus on the gapped phase and we derive the low- and high-energy effective theory at high-order in perturbation. This allows us to show that 
{\em i)} the low-energy effective Hamiltonian is an extended toric code Hamiltonian with still static but {\em interacting Abelian anyons} ;   
{\em ii)} the high-energy excitations are free fermions, composed of a hardcore boson with an
attached string of spin operators, hopping on a square lattice embedded in a magnetic field.

Our approach has some advantages which compensate for its perturbative nature and its restriction to the gapped phase. It provides a unified picture of the emergence of high- and low-energy excitations without introducing fermionic fields by hand as is done with Majorana or Jordan-Wigner fermionization. 
In addition, we work, from the beginning, in the thermodynamical limit, and derive analytical results for non-translational-invariant low-energy states. 


Let us consider the model in the limit $J_x,J_y\ll J_z$. In the limiting case
$J_x=J_y=0$ the model is a collection of isolated $z$-dimers. Each dimer has
four possible configurations : two low-energy states 
$\{|\!\uparrow\uparrow\rangle,|\!\downarrow\downarrow\rangle\}$ with energy
$-J_z$ and two high-energy states
$\{|\!\uparrow\downarrow\rangle,|\!\downarrow\uparrow\rangle\}$ with energy
$J_z$.
It is thus natural to interpret the change from a ferromagnetic to an
antiferromagnetic dimer configuration as the creation of a particle, with an
energy cost that we set equal to $1$ by chosing $J_z=1/2$. By construction,
these particles are hardcore bosons hopping on the sites of an effective
square lattice (see Fig.~\ref{fig:mapping_brickwall_square}), together with an
effective spin-$1/2$ indicating which kind of (anti-)ferro dimer configuration
is realized. Among the four possible mappings we choose the following
%
%
\begin{equation}
  \label{eq:mappings}
  |\!\uparrow\uparrow\rangle=|\!\Uparrow\! 0\rangle,
  |\!\downarrow\downarrow\rangle=|\!\Downarrow\! 0\rangle,\,
  |\!\uparrow\downarrow\rangle=|\!\Uparrow\! 1\rangle,
  |\!\downarrow\uparrow\rangle=|\!\Downarrow\! 1\rangle,
\end{equation}
%
%
where the left (right) spin is the one of the black (white) site of the dimer,
and double arrows represent the state of the effective spin. 
Let us denote by $\crb_\bi$ ($\anb_\bi$) the creation (annihilation) operator
of a hardcore boson at site $\bi$ (bold letters denote effective sites),
and $\tau_\bi^\alpha$ the Pauli matrices of the effective spin at the same
site. With these notations, the number of bosons in the system is
$Q=\sum_\bi\crb_\bi\anb_\bi$ and the Hamiltonian (\ref{eq:ham}) can be rewritten
as \cite{Vidal08_2}
%
%
\begin{equation}  \label{eq:ham_v2}
  H = -\frac{N}{2}+Q+T_0+T_{+2} + T_{-2},
 \end{equation}
%
%
where $N$ is the number of $z$-dimers,
%
%
\begin{eqnarray}
 T_0&=&-\sum_\bi \left(J_x\, t_\bi^{\bi+\bn_1}+J_y\, t_\bi^{\bi+\bn_2}
    +\mathrm{h.c.}\right),\\
  T_{+2}&=&-\sum_\bi \left(J_x\, v_\bi^{\bi+\bn_1}+J_y\, v_\bi^{\bi+\bn_2}\right)
  =T_{-2}^\dagger,
\end{eqnarray}
%
%
with hopping and pair creation operators
%
%
\begin{eqnarray}
  \label{eq:t}
  t_\bi^{\bi+\bn_1}=\crb_{\bi+\bn_1}\anb_\bi\, \tau^x_{\bi+\bn_1},\quad
  t_\bi^{\bi+\bn_2}=-\rmi\, \crb_{\bi+\bn_2}\anb_\bi\,
  \tau^y_{\bi+\bn_2}\tau^z_\bi , &&\\
  \label{eq:v}
  v_\bi^{\bi+\bn_1}=\crb_{\bi+\bn_1}\crb_\bi\, \tau^x_{\bi+\bn_1},\quad
  v_\bi^{\bi+\bn_2}=\rmi\, \crb_{\bi+\bn_2}\crb_\bi\,
  \tau^y_{\bi+\bn_2}\tau^z_\bi.&&
\end{eqnarray}
%
%
The vectors $\bn_1$ and $\bn_2$ are shown in Fig.~\ref{fig:mapping_brickwall_square}b.
Let us underline that  $[W_p,t_\bi^{\bj}]=[W_p,v_\bi^{\bj}]=0$ for all $(p,\bi,\bj)$ where the conserved plaquette operators \cite{Kitaev06} read (with notations given in Fig.~\ref{fig:mapping_brickwall_square}) 
%
%
\begin{equation}
W_p=\sigma_1^x\sigma_2^y\sigma_3^z\sigma_4^x\sigma_5^y\sigma_6^z=(-1)^{\crb_{\mbox{\tiny L}}\anb_{\mbox{\tiny L}}
  +\crb_{\mbox{\tiny D}}\anb_{\mbox{\tiny D}}}
\tau_{\mbox{\tiny L}}^y\,\tau_{\mbox{\tiny U}}^z\,
\tau_{\mbox{\tiny R}}^y\,\tau_{\mbox{\tiny D}}^z.
\end{equation}
%
%

At this stage, note that both the mapping (\ref{eq:mappings}) and the form of the Hamiltonian (\ref{eq:ham_v2}) are simply an alternative description of the problem which is always valid, even in a nonperturbative regime. The main difficulty resides in the fact that, now,  one has to deal with hardcore bosons coupled to effective spin degrees of freedom. Of course, one could use a fermionization trick and solve the model directly as done by Kitaev \cite{Kitaev06}. However, the procedure used in the following can be applied to nonexactly solvable models and further allows, in the present case, for a clear identification of the excitations. 

In general, a Hamiltonian of the form (\ref{eq:ham_v2}) cannot be diagonalized
exactly. Here, following Kitaev, we choose to treat it
perturbatively in the limit $J_x,J_y \ll J_z$. Nevertheless, the Green's functions
method used in Ref.~\onlinecite{Kitaev06} turns out to be rather hard to
implement at high order and/or high energy. Instead, we use an alternative
approach based on continuous unitary transformations (CUTs) \cite{Wegner94} whose
perturbative version \cite{Stein97,Knetter00} is especially well-suited to the problem at hand. Technical details will be given in a forthcoming
publication \cite{Vidal08_2}.

The main idea is to transform the Hamiltonian (\ref{eq:ham_v2}) which does not conserve the number of bosons into an effective Hamiltonian $H_\mathrm{eff}$ which satifies $[H_\mathrm{eff},Q]=0$. As explained in Ref.~\onlinecite{Knetter03}, $H_\mathrm{eff}$
is a sum of $k$-quasi-particle (QP) operators with $k\in \mathbb{N}$. 
The $k=0,1$ contributions can be written as
%
%
\begin{eqnarray} \label{eq:Heff}
  H_\mathrm{eff}^{0\,\mathrm{qp}} &=& E_0
  -\sum_{\{p_1,\ldots,p_n\}} C_{p_1,\ldots ,p_n} W_{p_1} W_{p_2}\ldots W_{p_n}\,,\hspace{0.2cm}\\
  H_\mathrm{eff}^{1\,\mathrm{qp}} &=& \mu Q
  -\sum_{\{\bj_1,\ldots,\bj_n\}}D_{\bj_1,\ldots ,\bj_n}
  t^{\bj_n}_{\bj_{n-1}}\ldots t^{\bj_3}_{\bj_2} t^{\bj_2}_{\bj_1}\,.
\end{eqnarray}
%
%
Here $\{p_1,p_2,\ldots,p_n\}$ denotes a set of $n$ plaquettes and the sum, $\{\bj_1,\ldots,\bj_n\}$ represents a sequence of $n$ connected sites.
The perturbative aspect  comes from the fact that the coefficients appearing in
$H_\mathrm{eff}$ are obtained as series expansions in $J_x$ and $J_y$.  We have
computed the 0-QP amplitudes $E_0$ and $C_{p_1,\ldots ,p_n}$ up to order
10 and the 1-QP amplitudes $\mu$ and $D_{\bj_1,\ldots,\bj_n}$ up to order
4. These expressions, being quite lengthy, will be given in a longer paper \cite{Vidal08_2}, as well as a discussion of ($k\geqslant 2$)-QP sectors.

\begin{figure}[ht]
  \includegraphics[height=7cm]{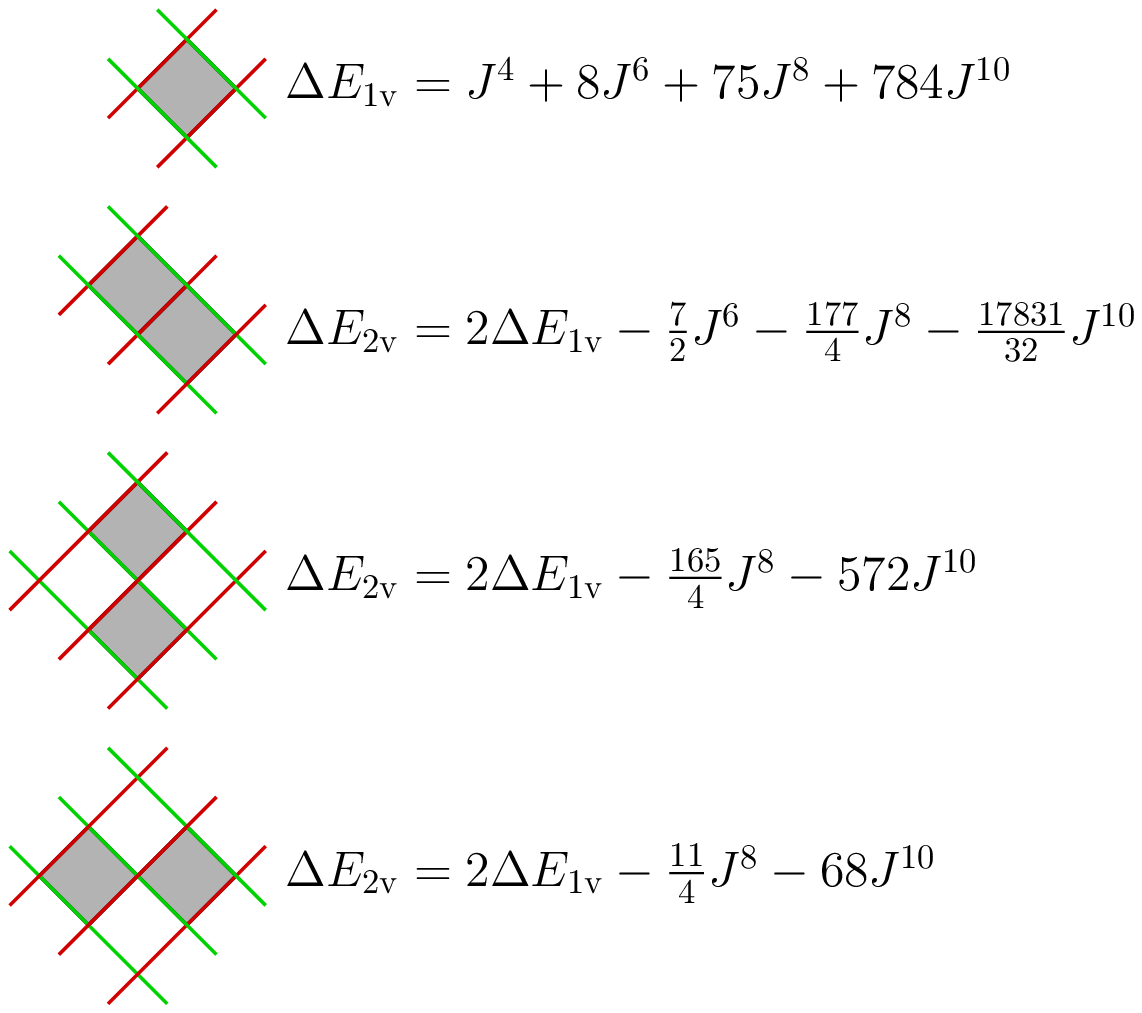}
  \caption{(color online). One-anyon and some two-anyon configurations
    (grey plaquettes) on a vortex-free background.
    $\Delta E_{1\mathrm{v}}$ ($\Delta E_{2\mathrm{v}}$) is the energy cost for
    adding one vortex (two vortices) to the vortex-free state. We have set
    $J_x=J_y=J$.}
  \label{fig:anyonic_energies}
\end{figure}

%
%
%
%
Let us first discuss the low-energy physics, i.e. the spectrum of 
$H_\mathrm{eff}^{0\,\mathrm{qp}}$. The most striking result is that the latter is only expressed in terms of the conserved quantities $W_p|_{0\,\mathrm{qp}}=\tau_{\mbox{\tiny L}}^y\,\tau_{\mbox{\tiny U}}^z\,
\tau_{\mbox{\tiny R}}^y\,\tau_{\mbox{\tiny D}}^z$. Thus, for any (vortex) configuration of the $W_p$'s which can take two values $\pm 1$, one readily gets the ground state energy of the corresponding sector. We wish to emphasize that at order 4 (lowest nontrivial order), as already discussed by Kitaev \cite{Kitaev06},  the only nonvanishing contribution (apart from $E_0$) involves only single plaquette-terms and one recovers the toric code Hamiltonian with its Abelian anyons \cite{Kitaev03}. 
At order 6, one gets the following corrections to these terms
%
%
\begin{eqnarray}
\frac{E_0}{N} &=&-\frac{1}{2}-\frac{J_x^2+J_y^2}{2}-\frac{J_x^4+J_y^4}{8} -\frac{J_x^6+J_y^6}{8}, \\
C_p&=&\frac{1}{2} J_x^2 J_y^2+\frac{J_x^4 J_y^2 +J_x^2 J_y^4}{4},   
\end{eqnarray}
%
%
but, more interestingly, one also has some two-plaquette terms :
%
%
\begin{equation} 
C_{p,p+\bn_1}=\frac{7}{8} J_x^4 J_y^2 \quad , \quad C_{p,p+\bn_2}=\frac{7}{8} J_x^2 J_y^4.
\end{equation} 
%
%
Actually, when increasing the perturbation theory order, one generates higher and higher multi-plaquette terms.  Thus, {\em the low-energy effective theory of the Kitaev model turns out to be an interacting anyon theory} whose eigenstates are those of the toric code. The
single-vortex energy and the two vortices interaction energies of three
configurations are given in Fig.~\ref{fig:anyonic_energies} at order 10, for
$J_x=J_y=J$.
%
%
%
%
It should finally be noticed that although the vortices interact, they remain static since $W_p$'s are conserved quantities.

\begin{figure}[ht]
  \includegraphics[height=5cm]{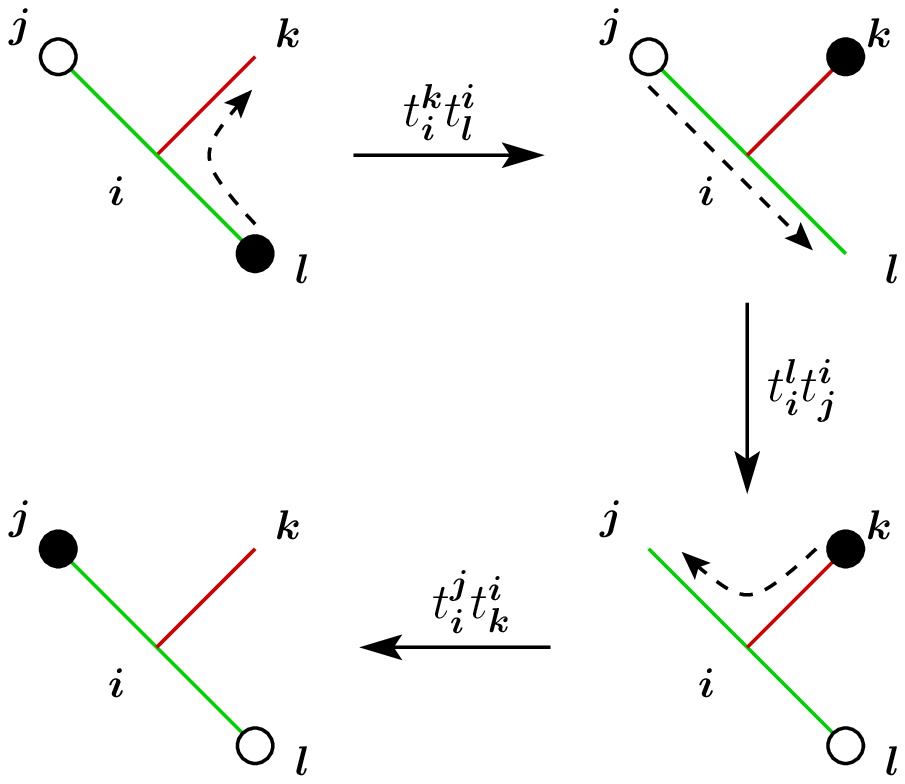}
  \caption{(color online). Illustration of the exchange of two particles discussed in the text,
    for $\bj=\bi+\bn_2$, $\bk=\bi+\bn_1$ and $\bl=\bi-\bn_2$.}
  \label{fig:exchange}
\end{figure}

In each sector defined by a configuration of the $W_p$'s, we shall now see that the excitation spectrum is of fermionic nature. As proposed by Levin and Wen \cite{Levin03}, the statistics can be directly determined from the hopping operators algebra. Let us consider the exchange process depicted in Fig.~\ref{fig:exchange}. The corresponding operator sequence is :
%
%
\begin{equation} 
t^{\bj}_{\bi}  t^{\bi}_{\bk}   t^{\bl}_{\bi}   t^{\bi}_{\bj}   t^{\bk}_{\bi}   t^{\bi}_{\bl}=-1 ,
\end{equation} 
%
%
or, equivalently,   $t^{\bi}_{\bj}   t^{\bk}_{\bi}   t^{\bi}_{\bl}=- t^{\bi}_{\bl}   t^{\bk}_{\bi}   t^{\bi}_{\bj}$. This latter identity shows that {\em the quasi-particles made of a hardcore boson and an effective spin-1/2 obey fermionic statistics}. 

%
%
%
%

We shall now see that solving the excitation spectrum is completely equivalent
to solving a problem of free fermions in a magnetic field on the square
lattice. Therefore, let us focus on the first-order perturbation theory for
which $H_\mathrm{eff}^{1\,\mathrm{qp}} = Q+T_0$ and consider a single
quasi-particle. Then, one can easily see that the momenta
$\big(H_\mathrm{eff}^{1\,\mathrm{qp}} -1 \big)^m$ for all $m$ are strictly equal
to those of a pure hopping fermion Hamiltonian in a magnetic field provided the
magnetic flux per plaquette mimicks the configuration of the $W_p$'s. This is
easily seen by noting that the product of the hopping operator around a
plaquette $p$ is given by $t^\bi_\bl t^\bl_\bk t^\bk_\bj t^\bj_\bi =W_p$,
with $\{\bi,\bj,\bk,\bl\}$ being any sequence of connected sites chosen among
L, U, R and D (see Fig.~\ref{fig:mapping_brickwall_square}).

\begin{figure}[ht]
  \includegraphics[height=4cm]{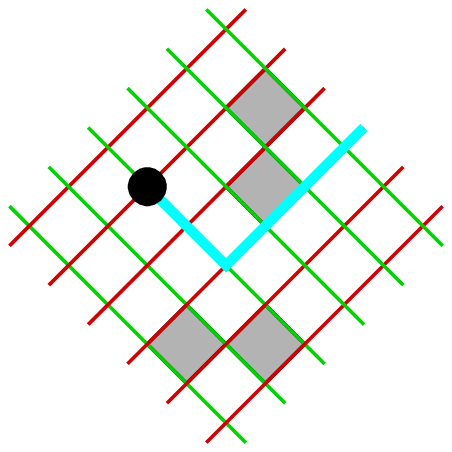}
  \caption{(color online). Pictorial representation of an emergent fermion as a composite object
    made of a hard-core boson (dot) and an attached string of spin-flips (thick
    line). Grey plaquettes are anyons, as in Fig.~\ref{fig:anyonic_energies}.}
  \label{fig:string}
\end{figure}

The $W_p$'s being conserved quantities, from a purely spectral point of view one
can replace the Pauli matrices in the hopping operators by pure numbers $\pm 1$
constraining the fluxes per plaquette (in unit of the flux quantum) to be
$\phi_p=0$ for $W_p=+1$ or $\phi_p=1/2$ for $W_p=-1$. At higher order, this
mapping remains true (with hoppings from one site to any other site) but one
needs to go beyond this simple argument to prove the correspondence between
both spectra \cite{Vidal08_2}. Let us simply mention that this is done by
building a basis of the 1-QP subspace, which turns out to 
be made of states with one string of spin flips attached to one
hard-core boson (yielding a fermion), as can be inferred from the form
(\ref{eq:t}) of the hopping operators, and illustrated in
Fig.~\ref{fig:string}. Note that the string fluctuates, as in the construction of anyons in \cite{Kitaev03}.

%
%
%
%

To illustrate the mapping, let us consider the 1-QP vortex free state
($W_p=+1, \forall p$). The spectrum of $H_\mathrm{eff}^{1\,\mathrm{qp}}$ is
made of a single band whose dispersion relation is given, at second order, by :
%
%
\begin{eqnarray} 
E_{\rm free}(k_x,k_y)&=&1-2[J_x \cos(k_x)+J_y \cos(k_y)] \\
&&+2[J_x \sin(k_x) + J_y \sin(k_y) ]^2,\nonumber
\end{eqnarray} 
%
%
where the wave vector $(k_x,k_y) \in [-\pi,\pi] \times [-\pi,\pi]$. The gap in this sector is then obtained by minimizing $E_{\rm free}$ and one gets $\Delta_{\rm free}=1-2(J_x+J_y)$. Note that, in this case, the perturbative result at order 1 coincides with the nonperturbative result obtained by Kitaev \cite{Kitaev06} and one recovers the transition point at $J_x+J_y=1/2=J_z$.

Let us complete our analysis with a comparison to the Majorana (or Jordan-Wigner) fermion
formalism. In the latter, one can compute the full spectrum exactly, by diagonalizing,  for each configuration of the $W_p$'s with $2N$ spins,  a $2N\times 2N$ matrix. Within our approach, we have a $N \times N$ matrix to consider.
Although we have seen that analytical (perturbative) results can be obtained
for $H_\mathrm{eff}^{0\,\mathrm{qp}}$, even for non-translational-invariant $W_p$'s, for
$H_\mathrm{eff}^{1\,\mathrm{qp}}$ the problem is thus essentially as complicated as in the Majorana (or Jordan-Wigner) fermion formalism. In this case,  the main interest of our method is thus to give a physical picture of the high-energy excitations as emerging fermions.

Furthermore, the continuous unitary transformations allow for the
calculation of correlation functions. In the 0-QP subspace, as we show below, some of these can again be expressed as a series in $W_p$ operators,
as was the case for the eigenvalues of $H_\mathrm{eff}^{0\,\mathrm{qp}}$. 
Here, we focus on the correlation function 
$C^{zz}_\bi=\langle\sigma^z_{\bi,\bullet}\sigma^z_{\bi,\circ}\rangle$
where $\bi,\bullet$ and $\bi,\circ$ are the black and white sites of a given
$z$-dimer $\bi$ (see Fig.~\ref{fig:mapping_brickwall_square}). Such a
correlation function can be computed exactly for a translationnaly invariant
configuration of $W_p$'s, either by using the technique developped in \cite{Baskaran07} or, in a much simpler way, thanks to the Hellman-Feynman
theorem \cite{Vidal08_2}. However, none of these two methods can be used effciently for  a
non-translational-invariant configuration of the $W_p$'s. 
%
%
\begin{figure}[t]
  \includegraphics[width=\columnwidth]{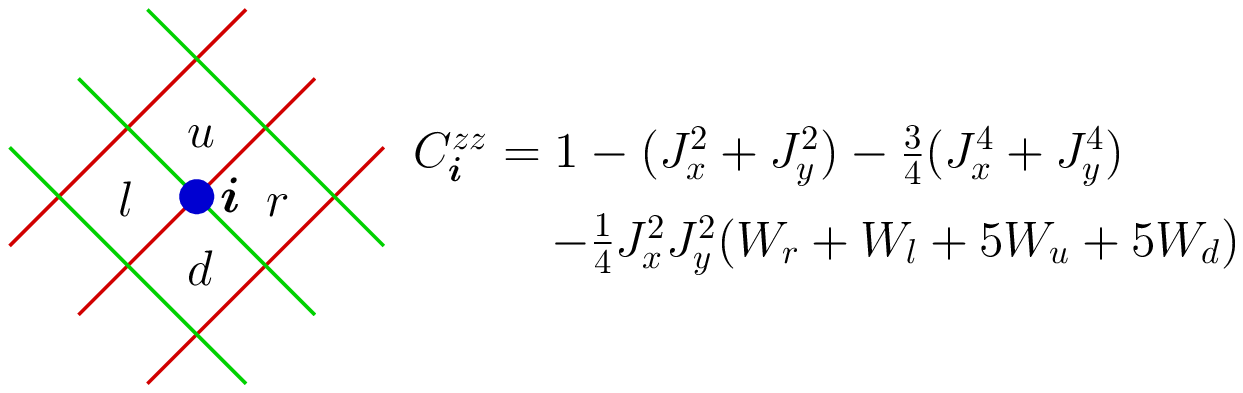}
  \caption{(color online). Perturbative expansion at order 4, of the correlation function
    $C^{zz}_\bi$ defined in the text, in the 0-QP subspace.}
  \label{fig:observable}
\end{figure}
%
%

As a first step, one has to translate $C^{zz}_\bi$ into the effective
spin and boson degrees of freedom, which yields
$C^{zz}_\bi=\langle(-1)^{\crb_\bi\anb_\bi}\rangle$. Then, the
observable appearing in the angular brackets has to be transformed with the same unitary
transformation as the one applied to the Hamiltonian \cite{Vidal08_2,Knetter03}. 
As an illustration, we give the result in Fig.~\ref{fig:observable} at order 4 which is the lowest nontrivial order where the $W_p$'s appear in the expansion.
Note that we have used the same notation $W$ for the operator and its expectation value on a 0-QP state.
This result shows how the correlation function is modified by the presence of
surrounding anyons. Of course, at higher order, anyons further apart would
also contribute \cite{Vidal08_2} as was already the case for the spectrum of $H_\mathrm{eff}^{0\,\mathrm{qp}}$.

To conclude, we wish to emphasize that the method we have used here, based on a perturbative approach of the CUTs, can be applied to nonexactly solvable models as well. In addition, as we have shown, it is especially efficient to compute the spectrum but also the expectation value of observables, even at high order in perturbation. Thus, we strongly hope that it constitutes a powerful tool to investigate, for instance, the non-Abelian anyons recently proposed  by Yao and Kivelson for a time-reversal symmetry breaking version of the Kitaev model \cite{Yao07, Vidal08_2}.

\begin{acknowledgments}
We thank B. Dou\c{c}ot for fruitful and stimulating discussions.
\end{acknowledgments}


\end{document}